\documentstyle[12pt,epsf]{article}
\catcode`\@=11

\def\section{\@startsection {section}{1}{\z@}{-3.0ex plus -1ex minus
    -.2ex}{0.5ex plus .2ex}{\bf }}
\def\subsection{\@startsection{subsection}{2}{\z@}{-1.5ex plus -1ex minus
   -.2ex}{0.3ex plus .2ex}{\it }}

\def\abstracts#1{{\centering{\begin{minipage}{13.0truecm}
        \footnotesize\baselineskip=12pt\noindent
        \parindent=0pt #1 \end{minipage}}\par}}

\renewenvironment{thebibliography}[1]
       {\vspace{-3.5ex} \begin{list}{\arabic{enumi}.}
        {\usecounter{enumi}\setlength{\parsep}{0pt}
        \footnotesize\baselineskip=12pt\noindent
        \setlength{\itemsep}{0pt} \settowidth
        {\labelwidth}{#1.}
        \settowidth{\leftmargin}{#1.\rule{3mm}{0mm}}
        \sloppy}}{\end{list}}

\newcommand{\fcaption}[1]{\vspace{1ex}
        \refstepcounter{figure}
        \setbox\@tempboxa = \hbox{\footnotesize {\bf Fig.~\thefigure.} #1}
        \ifdim \wd\@tempboxa > 14cm
           {\begin{center}
        \parbox{14cm}{\footnotesize\baselineskip=12pt {\bf Fig.~\thefigure.} #1}
            \end{center}}
        \else
             {\begin{center}
             {\footnotesize {\bf Fig.~\thefigure.} #1}

              \end{center}}
        \fi}

\newcommand{\tcaption}[1]{
        \refstepcounter{table}
        \setbox\@tempboxa = \hbox{\footnotesize {\bf Table~\thetable.} #1}
        \ifdim \wd\@tempboxa > 14cm
           {\begin{center}
        \parbox{14cm}{\footnotesize\baselineskip=12pt {\bf Table~\thetable.} #1}
            \end{center}}
        \else
             {\begin{center}
             {\footnotesize {\bf Table~\thetable.} #1}
              \end{center}}
        \fi
        \vspace{1ex}}

\def\@citex[#1]#2{\if@filesw\immediate\write\@auxout
        {\string\citation{#2}}\fi
\def\@citea{}\@cite{\@for\@citeb:=#2\do
        {\@citea\def\@citea{,}
        {\csname b@\@citeb\endcsname}}}{#1}}
\def\@cite#1#2{$\null^{#1}$}

\begin{document}

\begin{center}
 {\bf   CUMULANT EXPANSION METHOD IN THE DENSITY
 MATRIX APPROACH TO WAVE PACKET DYNAMICS IN MOLECULAR SYSTEMS
}\\
 \vspace{0.5cm} {\footnotesize
        Michael Schreiber\\
 {\it   Institute fur Physik, Techniche Univesitat, D-09107 Chemnitz, Germany
 }       \\
 {\tt   schreiber@physik.tu-chemnitz.de}\\
 \vspace{0.2cm}
        Dmitry Kilin\\
 {\it   Belarus State University, F.Scarina Avenue 4, Minsk, Belarus
    }}    \\
\end{center}

\vspace{0.3cm}
\abstracts{A non-Markovian master equation for a molecule interacting with a heat bath
is obtained using the cumulant expansion method. The equation
is applied to the problem of molecular wave packet relaxtional dynamics.
An exact solution is derived that demonstrates classical type of squeezing for
the wave packet evolution and a heat-bath dependent frequency shift.}
\section{ Introduction and model Hamiltonian}
The problem of relaxational dynamics of wave packets in molecular systems
is one of the up-to-date
physical problems supported by femtosecond-time-scale
spectroscopic facilities\cite{1} as well as by single molecule
spectroscopy.\cite{2}  Numerous theoretical investigations of the
problem appeal to different approaches.\cite{3}  Due
to the complexity of molecular relaxational dynamics a derivation of an
exactly solving model for the problem has not been obtained yet.
 In this paper we
investigate one of the models based on a master equation approach.

We will start from the general potential for a single molecule
interacting with a bath of harmonic oscillators. After the derivation of the
general master equation we will proceed to the specific example of harmonic
potential for the model, which admits an exact solution.

The molecule interacting with a heat bath is separated into relevant
 system
of diabatic vibronic levels $E_\mu$ and the environment of the
 heat bath modes
$\xi$ with frequency $\omega_\xi$,
 and creation operator $b^{+}_{\xi}$.
It is modelled by the Hamiltonian $H = H^S+H^E+H^{SE}$, with
\begin{equation}
\begin{array}{lcl}
H^S& =&\sum\nolimits_\mu E_\mu d_{\mu \mu }+\hbar \sum\nolimits_{\mu \nu }
v_{\mu\nu }d_{\mu \nu }  \\
H^E& =&\sum\nolimits_\xi \hbar \omega _\xi \left( b_\xi ^{+}b_\xi +
1/2\right),    \\
H^{SE}& =&\sum\nolimits_{\mu \nu }\hbar\left( r_{\mu \nu }+r_{\mu \nu}^{+}
\right) d_{\mu \nu },
\end{array}
\end{equation}
where $d_{\mu \nu }=\left| \mu \right\rangle \left\langle \nu \right| $ is
the transition operator of the system; $r_{\mu \nu }=\sum\nolimits_\xi {\cal K}_{\mu
\nu }^\xi b_\xi $ is annihilation operator of the bath
including the matrix elements
\noindent$ K%
_{\mu \nu }^\xi $ of interaction function $K^{\xi}$.
This model was discussed previously\cite{3} and an equation of
motion for the reduced density matrix $\sigma$ was derived taking
the system-environment coupling into account by perturbation theory.
In our approach to the
reduced density matrix equation we use the cumulant expansion method.\cite{4}
The applicability of this method is rather wide because it does not
appeals to
perturbation theories motivation but relates to the statistical aspects of the
influences of the heat bath.
\section{Master equation and its reduced form for the harmonic potential}
The second order cumulant expansion\cite{4} gives\cite{5}
\begin{equation}
\begin{array}{rll}
\dot{\sigma} &=&-\frac i\hbar [ H^S,\sigma] +\exp{( -\frac
i\hbar H^St)} {\bf \dot{K}} exp{( \frac i\hbar
H^St)} \sigma,    \\
{\bf \dot{K}}\sigma  &=&( -\frac i\hbar ) ^2\int\nolimits_0^td\tau
\left\langle [ \tilde{H}^{ES}( t) ,[ \tilde{H}
^{ES}( \tau) ,\sigma ]] \right\rangle,
\end{array}
\end{equation}
\noindent where angle brackets mean averaging over environment and
$\tilde{H}^{ES}$ is the Hamiltonian in interaction picture. This approach
allows us to describe non-Markovian processes, when such factors as
$\exp{i\omega _{k\lambda }\left( \tau -t\right) }$
(where $\omega_{k \lambda}=H^S_{k \lambda}/\hbar$) contain memory effects.
We obtain for the matrix elements\cite{5}
$$
\begin{array}{ll}
{\bf \dot{\sigma}}_{\mu \nu } =-\frac i\hbar [ H_0^S,{\bf \sigma }] _{\mu \nu }
-\int\limits_0^td\tau \sum\limits_{k\lambda }
&[(\left\langle r_{\mu\kappa}(t) r_{\kappa\lambda }^{+}( \tau) \right\rangle +
\left\langle r_{\mu\kappa}^{+}( t) r_{\kappa\lambda }( \tau) \right\rangle )
e^{i\omega _{k\lambda}( \tau -t ) }{\bf \sigma }_{\lambda \nu }\\
&-( \left\langle r_{\lambda \nu }( \tau ) r_{\mu \kappa}^{+}
( t) \right\rangle +\left\langle r_{\lambda \nu }^{+}
(\tau) r_{\mu \kappa }( t) \right\rangle )
{\bf\sigma }_{\kappa \lambda }e^{i\omega _{\lambda \nu}( \tau -t)} \\
&-(\left\langle r_{\lambda \nu }(t) r_{\mu \kappa}^{+}
( \tau) \right\rangle +\left\langle r_{\lambda \nu}^{+}
( t) r_{\mu \kappa }( \tau) \right\rangle)
 e^{i\omega _{\mu \lambda }( \tau -t)}{\bf \sigma }_{\kappa\lambda } \\
&+(\left\langle r_{\kappa \lambda }( \tau ) r_{\lambda \nu}^{+}
( t) \right\rangle +\left\langle r_{\kappa\lambda }^{+}
(\tau) r_{\lambda \nu }(t) \right\rangle )
{\bf\sigma }_{\mu \kappa }e^{i\omega _{k\lambda }( \tau -t)} ].
\end{array}
$$
Using the Markov approximation the master equation reads\cite{5}
$$
\begin{array}{ll}
{\bf \dot{\sigma}}_{\mu \nu }{=}-\frac i\hbar \left[ H_0^S,{\bf \sigma }
\right] _{\mu \nu }{-}\pi \sum\limits_{\kappa\lambda }\sum\limits_{\xi}
&[{\cal K}_{\mu k}^\xi {\cal K}_{k\lambda }^\xi \left\{ \left( 1{+}n _\xi  \right)
\delta \left( \omega _{k\lambda }{+}\omega _\xi
\right) {+}n_\xi \delta \left( \omega _{k\lambda}
{-}\omega _\xi \right) \right\} {\bf \sigma }_{\lambda \nu } \\
&-{\cal K}_{\lambda \nu }^\xi {\cal K}_{\mu \kappa }^\xi
\left\{ \left(1{+}n _\xi  \right) \delta \left( \omega _{\lambda\nu}
{-}\omega _\xi \right) {+}n_\xi \delta \left( \omega
_{\lambda v}{+}\omega _\xi \right) \right\} {\bf \sigma }_{\kappa \lambda } \\
&-{\cal K}_{\lambda \nu }^\xi {\cal K}_{\mu \kappa }^\xi
\left\{ \left(1{+}n_\xi \right) \delta \left( \omega _{\mu \lambda}{+}\omega _\xi \right) {+}n _\xi \delta \left( \omega
_{\mu \lambda }{-}\omega _\xi \right) \right\} {\bf \sigma }_{k\lambda } \\
&+{\cal K}_{\kappa \lambda }^\xi {\cal K}_{\lambda \nu }^\xi
\left\{ \left(1{+}n_\xi  \right) \delta \left( \omega _{k\lambda}
{-}\omega _\xi \right) {+}n_\xi \delta \left( \omega_{k\lambda }{+}
\omega _\xi \right) \right\} {\bf \sigma }_{\mu \kappa }].
\end{array}
$$
In the following we apply this approach to a system with a harmonic
oscillator potential. In the interacrion picture
the respective Hamiltonian of interaction reads
\begin{equation}
\begin{array}{rcl}
\tilde{H}^{ES}\left( t\right)& =&\sum\nolimits_\xi K^\xi \left( b_\xi ^{+}\left(
t\right) +b_\xi \left( t\right) \right) \left( a^{+}\left( t\right) +a\left(
t\right) \right).
\end{array}
\end{equation}
We also suggested ,
that there are no negative frequencies in bath. Returning to the Schr\"{o}dinger
picture we can finally write the master equation for a damped harmonic
oscillator \cite{5}
\begin{equation}
\begin{array}{ll}
\dot{\sigma}=-i\omega \left[ a^{+}a, \sigma \right]&+\gamma n \left( \left[ \left( a^{+}+a\right) ,\sigma
a\right] +\left[ a^{+}\sigma ,\left( a^{+}+a\right) \right] \right)  \\
&+\gamma ( n+1) \left( \left[ \left( a^{+}+a\right) ,\sigma
a^{+}\right] +\left[ a\sigma ,\left( a^{+}+a\right) \right] \right),
\end{array}
\end{equation}
with damping rate $\gamma n=\pi \left( K^\xi \right) ^2\rho
\left( \omega \right) n\left( \omega \right) $, $\rho$ means density of bath states. It should be noted
that this master equation differs from the usual form of the master equation for
the damped harmonic oscillator\cite{6}. The differences are the result
of the full form of the molecule-bath interaction. Usually rotating wave approximation
form of this Hamiltonian is used. The presence of the counter-rotating terms\cite{5}
in the Hamiltonian leads to the phase-dependent terms in the master equation
which connect diagonal matrix elements $\sigma _{nn}$
and non-diagonal ones $\sigma _{mn}$. The later describes the phase-dependent
effects.

\section{Method of solution and analytical results}

The solution is based on an application of the characteristic function $F\left( \lambda
,\lambda ^{*},t\right) =Sp\left( f\sigma \right) $, where $f=e^{\lambda
a^{+}}e^{-\lambda ^{*}a}.$ Evaluating as usually commutators
like $\left[ a,\exp{(\lambda a^{+})}\right] $ one finds\cite{5}: $Sp(\left[ a^{+}a,\sigma \right] f)
=\left( \lambda ^{*}\partial _{\lambda
^{*}}-\lambda \partial _\lambda \right) F$ , $Sp(\left[ a^{+},\sigma a\right] f)
=\lambda \left( \partial _{\lambda
^{*}}-\lambda \right) F$ , $Sp(\left[ a^{+}, \sigma a^{+}\right] f)=-\lambda
^{*}\partial _\lambda F$, and etc. This allows us to transfer the operator equation (4) into the c-number
differential equation with partial derivatives
\begin{equation}
\dot{F}= \left( -\left( i\omega \lambda ^{*}+\gamma \left( \lambda
^{*}+\lambda \right) \right) \partial _{\lambda ^{*}}+\left( i\omega \lambda
-\gamma \left( \lambda ^{*}+\lambda \right) \right) \partial _\lambda
-\gamma n\left\{ \lambda +\lambda ^{*}\right\} ^2\right) F.
\label{nine}
\end{equation}
The solution of the equation is obvious. We solve it by expanding $F\left( \lambda ,\lambda ^{*},t\right) =\exp
\left( \sum_{m,n}K_{mn}\left( t\right) \lambda ^m\left( -\lambda ^{*}\right)
^n\right) $. Substitution $F$ in this form into (\ref{nine}) gives a set of independent systems of differential equations (SODE)\cite{5}
for functions $K_{mn}$ with $m+n=N$ being a fixed number. The first two systems are:\\ \\$
\begin{array}{cc}
\begin{array}{c}
\dot{K}_{10}=\left( i\omega -\gamma \right) K_{10}+\gamma K_{01}, \\
\noindent
\dot{K}_{01}=\gamma K_{10}+\left( -i\omega -\gamma \right) K_{01},
\end{array}
&
\begin{array}{c}
\dot{K}_{11}=2\gamma \left( n-K_{11}\right) +2\gamma \left(K_{02}+K_{20}\right),  \\
\dot{K}_{20}=-\gamma \left( n-K_{11}\right) +2\left( i\omega -\gamma\right) K_{20}, \\
\dot{K}_{02}=-\gamma \left( n-K_{11}\right) -2\left( i\omega +\gamma\right) K_{02}.
\end{array}
\noindent
\noindent
\end{array}
\\
\newline
$For a wide class of initial states (coherent, thermal,
squeezed, etc.) the consideration can be restricted by these two
systems, because all higher-order elements of $K_{mn}$ are zero. Below we
choose the coherent state as initial one. For this initial state
the solutions of these SODEs read
\begin{equation}
\begin{array}{rcl}
K_{10}+K_{01}&=&Q_{0}e^{-\gamma t}\left[(\gamma/\tilde\omega)
\sin {\tilde\omega t}+\cos{ \tilde\omega t}\right],\\
K_{11}+K_{20}+K_{02}&=&n-ne^{-2\gamma t}\left[ 1+(\gamma/\tilde\omega)^2
 ( 1-\cos{2\tilde\omega t})
 + (\gamma/\tilde\omega)\sin{2\tilde\omega t}\right],
\end{array}
\end{equation}
where $\tilde\omega=\sqrt{\omega^2-\gamma^2}$, $n= ( \exp{(\hbar \omega /k_BT)} -1 ) ^{-1}$ and
$Q_{0}$ is initial coherent state coordinate. The distribution of
interest in coordinate space is ${P}\left( Q,t\right)$=\\
$\frac 1{2\pi }\int\nolimits_{-\infty }^\infty
d\lambda e^{-i\lambda Q}{\cal \chi }\left( \lambda,t \right)$,
where ${\cal \chi }\left( \lambda,t \right) =Sp\left( e^{i\lambda \left(
a^{+}+a\right) }\sigma(t)\right)$ =$e^{-\frac 12\lambda ^2}F\left( i\lambda ,-i\lambda,t
\right) $. Integration yields
\begin{equation}
P( Q,t) =\frac 2{\sqrt{\pi \left( \frac
12+K_{20}+K_{02}+K_{11}\right) }}
\exp \left\{ -\frac{\left( Q-K_{10}-K_{01} \right) ^2}{ \frac
12+K_{20}+K_{02}+K_{11} }\right\}.
\end{equation}
\newpage
The obtained exact solution $P(Q,t)$ shows some kind of classical squeezing
%\vspace{8cm}
\begin{figure} \begin{center}
%\fbox{\epsfysize=8cm\epsfbox{pl.ps}}
%  \fbox{\epsfxsize=8cm}
%
\vspace{8cm}
% use the second line if you do not want to include postscript figures
  \fcaption{
The plots show wave packet dynamics $\gamma=0.1 \omega$,
$k_B T=3 \hbar \omega$, $\omega t=(0:20)$, $Q_0=4$ (a); and the variance of
the wave packet (b). }  \label{welc}
\end{center}\end{figure}
because the width of the wave packet slightly oscillates in time instead of it's
monotone increasing for usual damping processes. This squeezing which is displayed
in Fig.1 has not been derived previously. It should be
stressed that this is not quantum squeezing because this width is never
smaller than the ground state width. Eq.6  demonstrates a decrease of
the effective harmonic oscillator frequency due to the phase-dependent
interaction with the bath. This prediction is also new.

The consideration of more complex potentials on the basis of
proposed generalized master equations
is now in progress.


\begin{thebibliography}{9}
\bibitem{5} A more detailed derivation will be published elsewere
\end{thebibliography}
\end{document}